\documentclass[prd,article,nofootinbib,twocolumn,preprintnumbers,superscriptaddress]{revtex4}
\usepackage{graphicx,amsfonts,color,comment,amsmath,hyperref,float,times}
\usepackage{amssymb}	

\usepackage{mathrsfs,amssymb}  
\usepackage{cancel}
\usepackage[normalem]{ulem}
\usepackage[dvipsnames]{xcolor}
\usepackage{tikz}

\newcommand{\be}{\begin{equation}}
\newcommand{\ee}{\end{equation}}
\newcommand{\bea}{\begin{eqnarray}}
\newcommand{\eea}{\end{eqnarray}}

\newcommand{\dd}{\; \mathrm{d}}

\begin{document}

\title{Terrestrial Effects on Dark Matter-Electron Scattering Experiments}

\author{Timon Emken}
\email[E-mail: ]{emken@cp3.sdu.dk}
\affiliation{CP3-Origins, University of Southern Denmark, Campusvej 55, DK-5230 Odense, Denmark}

\author{Chris Kouvaris}
\email[E-mail: ]{kouvaris@cp3.sdu.dk}
\affiliation{CP3-Origins, University of Southern Denmark, Campusvej 55, DK-5230 Odense, Denmark}

\author{Ian M. Shoemaker}
\email[E-mail: ]{ian.shoemaker@usd.edu}
\affiliation{University of South Dakota, Department of Physics, 414E Clark St., Vermillion, SD 57069, USA}

\date{\today}
\begin{abstract}
A well-studied possibility is that dark matter may reside in a sector secluded from the Standard Model, except for the so-called photon portal: kinetic mixing between the ordinary and dark photons. Such interactions can be probed at dark matter direct detection experiments, and new experimental techniques involving detection of dark matter-electron scattering offer new sensitivity to sub-GeV dark matter. Typically however it is implicitly assumed that the dark matter is not altered as it traverses the Earth to arrive at the detector. In this paper we study in detail the effects of terrestrial stopping on dark photon models of dark matter, and find that they significantly reduce the sensitivity of XENON10 and DAMIC. In particular we find that XENON10 only excludes masses in the range (5-3000) MeV while  DAMIC only probes (20-50) MeV. 
Their corresponding cross section sensitivity is reduced to a window of cross sections between $(5\times 10^{-38}-10^{-30})~{\rm cm}^{2}$ for XENON10 and a small window around $\sim 10^{-31}~{\rm cm}^{2}$ for DAMIC. We also examine implications for a future DAMIC run.\\[0.2cm]
\textit{Preprint: CP3-Origins-2017-007 DNRF90}

\end{abstract}
\preprint{}


\maketitle

\section{Introduction}

Dark matter (DM) appears to be rather secluded from the visible sector. Indeed, to date only its gravitational impact on visible matter has been verified. There are a number of well-motivated models however in which a small coupling (or portal) to the dark sector is expected. A simple possibility is the so-called ``photon portal'' in which the ordinary SM photon kinetically mixes with a dark photon~\cite{Holdom:1985ag} which has been widely studied~\cite{Foot:2004pa,Feldman:2006wd,ArkaniHamed:2008qn,Pospelov:2008jd,Bjorken:2009mm}.
 ~This possibility is encapsulated in the Lagrangian
\be \mathscr{L} \supset g_{X} \bar{X} \gamma^{\mu} X A'_{\mu} + \varepsilon F_{\mu \nu}F'^{\mu \nu} + m_{\phi}^{2} A'_{\mu}A'^{\mu}
\label{eq:1}
\ee
where $X$ is the DM, $\gamma'$ is a dark photon with field strength $F'^{\mu \nu}$ and mass $m_{\phi}$, and $g_{X}$ is the new gauge coupling. It is important to notice that in such models the DM couples equally to the electrons and protons of the visible sector but not the neutrons. As a result this model can be searched for at traditional direct detection experiment~\cite{Goodman:1984dc} looking for nuclear recoil as well as the more recently proposed experiments with sensitivity to DM-electron scattering~\cite{Essig:2011nj,Essig:2012yx,Lee:2015qva,Essig:2015cda}.

In fact with XENON10 data~\cite{Angle:2011th} the first direct limits on sub-GeV dark matter from electron scattering have been derived~\cite{Essig:2012yx}. In this paper we re-visit these limits and re-interpret them in the context of the dark photon model after carefully considering the impact of the Earth overburden on the incoming DM. At the Laboratori Nazionali del Gran Sasso site the XENON10 detector was situated with $\sim 1400$ m of rock, so as to reduce the cosmic ray background. However given the relatively large DM cross sections being probed in this experiment, the rock overburden can additionally affect the DM as well. We will also apply our results to the DAMIC bounds discussed in~\cite{Essig:2015cda} derived from an engineering run of DAMIC~\cite{Barreto:2011zu}. Additional possible methods for detecting sub-GeV DM include superconductors~\cite{Hochberg:2015pha,Hochberg:2015fth},  superfluid helium~\cite{Schutz:2016tid,Knapen:2016cue}, as well as a search tactic employing conventional DM detectors wherein a detectable photon is emitted from the scattered nucleus~\cite{Kouvaris:2016afs}. 

In a previous paper~\cite{Kouvaris:2014lpa}, the authors examined the impact on DM deceleration from: (i) electronic Coulomb interactions in insulators, (ii) electronic Coulomb interactions in metals, and (iii) nuclear recoil. There they found however the nuclear stopping dominated over the other contributions. We shall therefore focus on nuclear stopping here. 

Although elastic nuclear recoils of sub-GeV DM might be undetectable in the laboratory, they could still have a critical effect on direct detection experiments based on inelastic DM-electron scatterings. Particularly in models like Eq.~\eqref{eq:1}, where the DM-nucleus scattering cross-section can be large, elastic nuclear recoils can modify the local DM velocity distribution by decelerating incoming DM particles before they enter the detector. This can be critical, especially since the experiments based on ionisation are typically sensitive to DM with high velocities. In the most extreme case, nuclear stopping in the Earth crust slows down DM particles to a degree, that ionisation in the detector can no longer occur. This way cross-sections above a certain critical value can no longer be probed at the experiment.

This was observed in earlier work~\cite{Lee:2015qva}, where the authors gave a rough estimate for the effect. In this work we present a more precise determination of the critical cross-section, at which nuclear stopping becomes sufficiently efficient that ionisation events in DM-electron scattering experiments are no longer possible, simply because the DM particles do not have enough energy. The limits based on this criteria are significantly stronger than those in~\cite{Lee:2015qva}. We present these results for XENON10 and DAMIC using both analytical and numerical methods. The latter are Monte Carlo simulations of DM particles scattering inside the Earth's crust. These allow to quantify the full effect of elastic DM-nucleus scatterings on the DM velocity distribution, which occur due to both energy loss and the deflection of the DM trajectory. Similar MC simulations have been used in \cite{Collar:1992qc,Collar:1993ss,Hasenbalg:1997hs}, where trajectories through the Earth are simulated to explore a diurnal signal modulation, and  in \cite{Zaharijas:2004jv} to describe the effect of nuclear stopping on direct detection experiments based on conventional nuclear recoil detection. Note that the diurnal modulation induced by underground scatterings of DM with atoms has been studied recently for generic DM-atom contact interactions in~\cite{Kavanagh:2016pyr}.

This paper is organized in the following manner. In the next section we review the analytic treatment of stopping presented in~\cite{Kouvaris:2014lpa}, and apply it to the dark photon model of Eq.~(\ref{eq:1}). In the next section we describe an improved treatment using Monte Carlo simulations to track DM through multiple scattering events before arriving at the detector depth.  


\section{Analytic Description of Nuclear Stopping}
\subsection{Review of Nuclear Stopping}
As a DM particle traverses the Earth it loses kinetic energy every time it scatters on a nucleus. This nuclear stopping induced energy loss can be described \cite{Starkman1990,Kouvaris:2014lpa} as a continuous process along the particle's path,
\be
\frac{\dd E}{\dd x} = - n_{N} \int_{0}^{E_{R}^{{\rm max}}} \frac{\dd \sigma}{\dd E_{R}} E_{R} \dd E_{R}\, ,
\label{eq:cont}
\ee
where $n_{N}$ is the number density of nucleus $N$ and 
\be 
\frac{\dd \sigma}{\dd E_{R}} = \frac{m_{N} \sigma_{N}}{2\mu_{N}^{2} v^{2}}=  \frac{m_{N} \sigma_{p}Z^{2}}{2\mu_{p}^{2}v^{2}}
\ee
is the DM-nucleus differential cross section, where the maximum nuclear recoil is $E_{R}^{{\rm max}} = \gamma E$ with $\gamma=\frac{4m_{X}m_{N}}{(m_{X}+m_N)^2}$. Throughout this work we consider sub-GeV DM particles, for which the nuclear form factor is approximately unity. We further assume that the mass scale of the gauge boson $A'$ is sufficiently heavy so that the scattering is effective contact and the DM form factor is, $F_{DM}(q)\approx 1$. Upon integration Eq.(\ref{eq:cont}) becomes 
\begin{align}
	dE/dx &= - \frac{n_{N}\gamma^{2} E^{2}}{2}\frac{\dd\sigma}{\dd E_{R}}\, ,\\
	\Rightarrow \log\left( \frac{E_{{\rm in}}}{E_{{\rm f}}}\right) &= \frac{2n_{N}  \sigma_{p} Z^{2} \mu_{N}^{4}L}{m_{X}m_{N} \mu_{p}^{2}}\, ,
\label{eq:cont2}
\end{align}
where $E_{\rm in}$ is the initial incoming DM kinetic energy and $E_{\rm f}$ is the energy after traversing a distance $L$ undergoing continuous elastic scatterings on nuclei of type $N$. We can solve this for the critical cross-section which will decelerate the DM to an $E_{\rm f}$ below the critical threshold energy for detection at a given experiment, $E_{\rm thr}$,
\be 
\sigma_{p}^{{\rm max}} \simeq \frac{m_{X}m_{N}\mu_{p}^{2}}{2n_{N} Z^{2} \mu_{N}^{4} L}  ~\log\left(\frac{E_{\rm i}}{E_{{\rm thr}}}\right)\, . \label{eq:sigmapmax}
\ee
This is the largest cross section that can be probed at an experiment at a depth $L$ with a threshold $E_{\rm thr}$. Note that the energy loss is assumed to happen along a straight path between the Earth surface and the detector. The fact that scatterings also cause deflections of the DM particles, which effectively prolong their paths between the surface and the detector, is not taken into account in this description, making Eq. \eqref{eq:sigmapmax} a conservative estimate.

\subsection{Critical cross-section for electron scattering experiments}
In the dark-photon model the cross-sections of DM-nucleus and DM-electron scatterings are related. In the heavy mediator case, this relation reads
\begin{align}
	\frac{\sigma_e}{\sigma_N} \simeq \left(\frac{\mu_e}{Z\mu_N}\right)^2\, ,\label{eq:eN}
\end{align}
where $\mu_e$ and $\mu_N$ are the reduced mass of the DM particle and the electron and nucleus respectively.

It becomes clear that DM-electron cross-sections being probed in the laboratory are accompanied by strong DM-nucleus interactions, potentially causing significant nuclear stopping in the Earth crust. Provided with such a relation, we find the maximum DM-electron scattering cross-section which can be probed, corresponding to Eq. \eqref{eq:sigmapmax}. Noting  $\frac{\sigma_{e}}{\sigma_{p}} \simeq  \left(\frac{\mu_{e}}{\mu_{p}}\right)^{2}$, we find
\be 
\sigma_{e}^{{\rm max}} \simeq \frac{m_{X}m_{N}\mu_{e}^{2}}{2n_{N} Z^{2} \mu_{N}^{4} L}  ~\log\left(\frac{E_{\rm i}}{E_{{\rm thr}}}\right).\label{eq:sigmaemax}
\ee

\begin{table}
	\centering
	\begin{tabular}{|l|l|l|}
	\hline
	Experiment	&Depth [m]	&$E_{\rm thr}$[eV]\\
	\hline
	\hline
	XENON10		&1400		&12.4\\
	DAMIC		&100			&40\\
	DAMIC (proj.)	&100		&$\sim 1-2$\\
	\hline
	\end{tabular}
	\caption{Here we summarize the relevant experimental parameters for XENON10 and DAMIC.}
	\label{tab:exp}
\end{table}

To obtain the cross-section, for which nuclear stopping spoils the possibility of detection, we have to make assumptions about the initial energy and the threshold. We assume that all incoming DM particles arrive with `optimal' conditions, i.e., that they arrive with maximum speed, $v_{\rm ini}=(v_{\text{esc}}+v_{\oplus})\approx 800\,\text{km}\,\text{s}^{-1}$, and point directly downwards towards the detector. The first advantage of this choice is that it is conservative. If the fastest particles can be screened off by the Earth's crust, then so can the other much slower particles. The second advantage is the independence of the DM halo model. The experiment-related specifics, the depth and energy threshold, are given in table \ref{tab:exp}.
\section{Monte-Carlo Simulation of Nuclear Stopping}
In the previous analytic treatment of nuclear stopping we did not take into account deflections and assumed that the deceleration of a DM particle travelling through the Earth crust occurs along a straight path. However, scattering deflections will increase the travelled distances and therefore the effect of nuclear stopping. To include deflections we set up a Monte Carlo simulation of particle trajectories in the Earth's crust.

\subsection{Scattering probability and kinematics}
 The mean free path of a particle moving through the Earth's crust is given by
\begin{align}
\lambda_{\text{MFP}}^{-1}&= \sum_{i}n_{N_i}\sigma^{\text{SI}}_{N_i}=\sum_{i}f_{N_i}\frac{\rho_{\text{crust}}}{m_{N_i}}\sigma^{\text{SI}}_{N_i}\, . \label{eq:lambdainverse}
\end{align}
\begin{table}[t!]
\centering
	\begin{tabular}{|l|l|}
		\hline
		Element		&Abund.[\%]	\\
		\hline
		\hline
		O			&46.6				\\
		Si			&27.7				\\
		Al			&8.1					\\
		Fe			&5.0					\\
		Ca			&3.6					\\
		K			&2.8					\\
		Na			&2.6					\\
		Mg			&2.1					\\
		\hline
		Total			&98.5			\\
		\hline
	\end{tabular}
	\caption{Approximate elemental abundances within the continental crust of the Earth~\cite{lutgens2003essentials}.}
	\label{tab:abundances}
\end{table}
The elemental abundances $f_{N_i}$ of nucleus species $i$ in the continental crust are given in table \ref{tab:abundances}. For the density we picked a value of $\rho_{\text{crust}}\approx 2.7\frac{\text{g}}{\text{cm}^3}$. The distance $L$ a particular particle travels without scattering on nuclei is given by
\begin{align}
	L=-\lambda_{\text{MFP}}\log\xi, \label{eq:L}
\end{align}
where $\xi$ is a uniformly distributed random number between 0 and 1. We then define the particle displacement vector $\overrightarrow{\Delta r}$ of magnitude $L$ and direction of $\vec{v}_X$. It is the vector which connects the particle's current position with the nucleus on which it scatters next,
\begin{align}
\overrightarrow{\Delta r} &=L \frac{\vec{v}_X}{|\vec{v}_X|}\, . \label{eq:freepathvector}
\end{align}	
After this displacement the DM particle scatters off a resting nucleus. The probability for this nucleus to be of species $i$ and mass $m_{N_i}$ is then
\begin{align}
	P_i=\frac{n_{N_i}\sigma^{\text{SI}}_{N_i}}{\sum_{j}n_{N_j}\sigma^{\text{SI}}_{N_j}}\, .\label{eq:P}
\end{align}
The DM particle's velocity after the scattering $(\vec{v}_{X})'$ is given by
\begin{align}
(\vec{v}_{X})' = \frac{1}{m_N+m_{X}}\left( m_N \left|\vec{v}_X\right| \vec{n}^{\text{CMS}}+m_{X}\vec{v}_X\right)\, .\label{eq:vnew}
\end{align}
The unit vector $\vec{n}^{\text{CMS}}$ points into the direction of the DM particle's velocity after the scattering in the center of mass frame. In the case of spin-independent DM-nucleus interactions the scattering is isotropic and $\vec{n}^{\rm CMS}$ points towards a uniformly distributed random direction.

\subsection{Algorithm}
\begin{figure}[b!]
\centering
\includegraphics[width=0.5\textwidth]{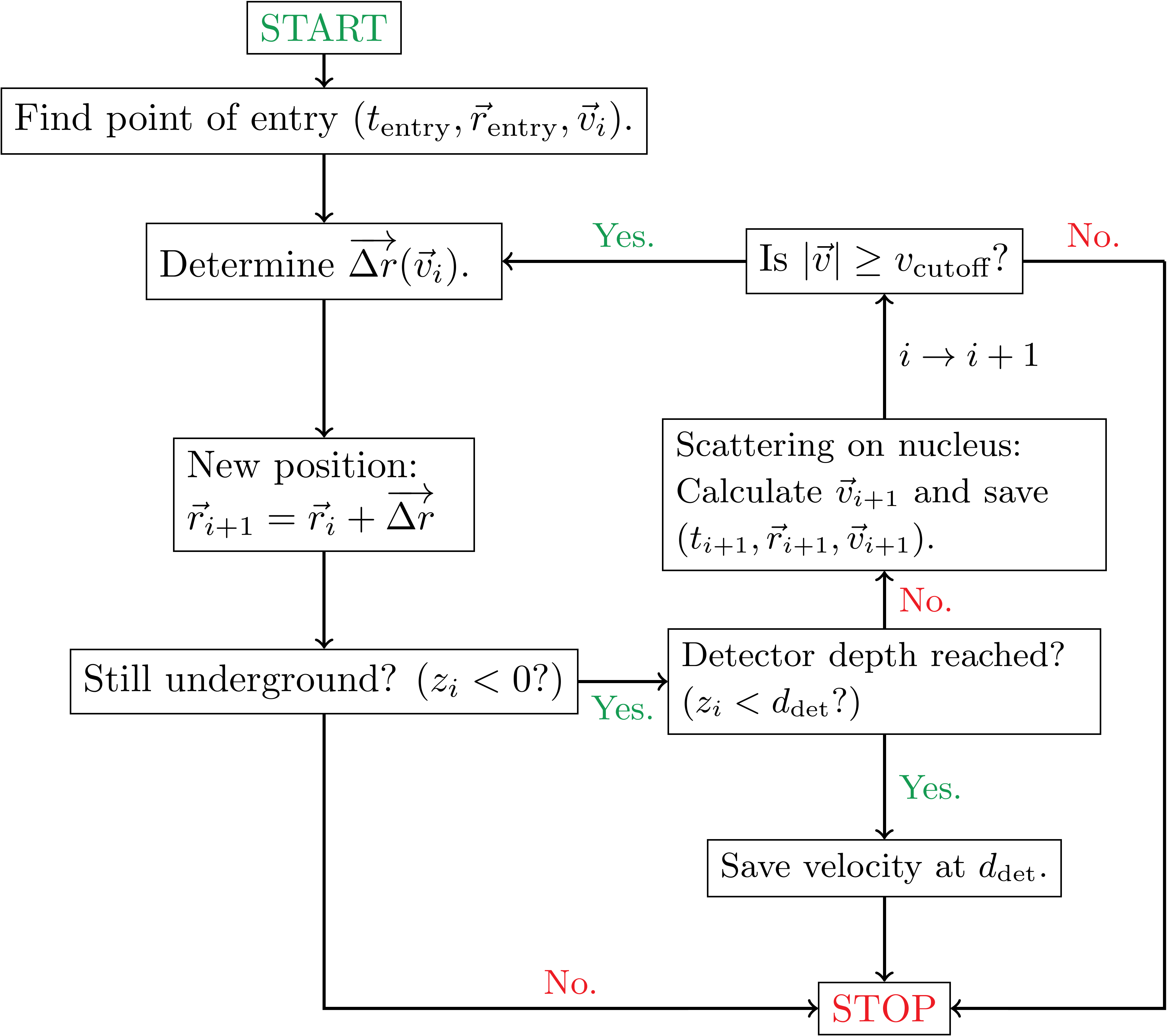}
\caption{Flowchart of the MC simulation algorithm.}
\label{fig:algorithm}
\end{figure}
We implement a random-walk algorithm to simulate the particles trajectory between the Earth surface and the detector underground depth while it gets deflected and decelerated by scatterings on nuclei of the Earth crust. Each simulated particle is followed, starting above the Earth surface with the same initial conditions as before, $\vec{v}_{\rm ini}\approx -800~ \text{km }\text{s}^{-1}\vec{e}_z$ ($\vec{e}_z$ being a unit vector pointing from the center of the Earth to the detector). The mean free path in the Earth crust is constant and Eqs. \eqref{eq:L} and \eqref{eq:freepathvector} provide us with the location of the first scattering event. Subsequently a sub-routine determines the nucleus species, which is involved in the scattering, according to Eq. \eqref{eq:P}. We then compute the DM particle's velocity after the scattering event via Eq. \eqref{eq:vnew}. After that the procedure repeats itself. The Eqs. \eqref{eq:L} and \eqref{eq:freepathvector} again give the location of the next scattering, and so on. The algorithm proceeds until one of three conditions is fulfilled.
\begin{enumerate}
	\item The particle gets deflected back outside the Earth.
	\item The particle reaches the detector's depth in the crust.
	\item The particle's speed falls below a cut-off speed. This is introduced for computational reasons and is chosen to be extremely low ($v_{\rm cutoff}\approx 10^{-10}\text{cm }\text{s}^{-1}$).
\end{enumerate}
If the second condition is fulfilled and the DM particle reaches the detector, we save its velocity. These particles will make up our velocity sample. A flowchart of the algorithm can be found in Fig. \ref{fig:algorithm}.

\begin{figure}[t!]
	\centering
	\includegraphics[width=0.5\textwidth]{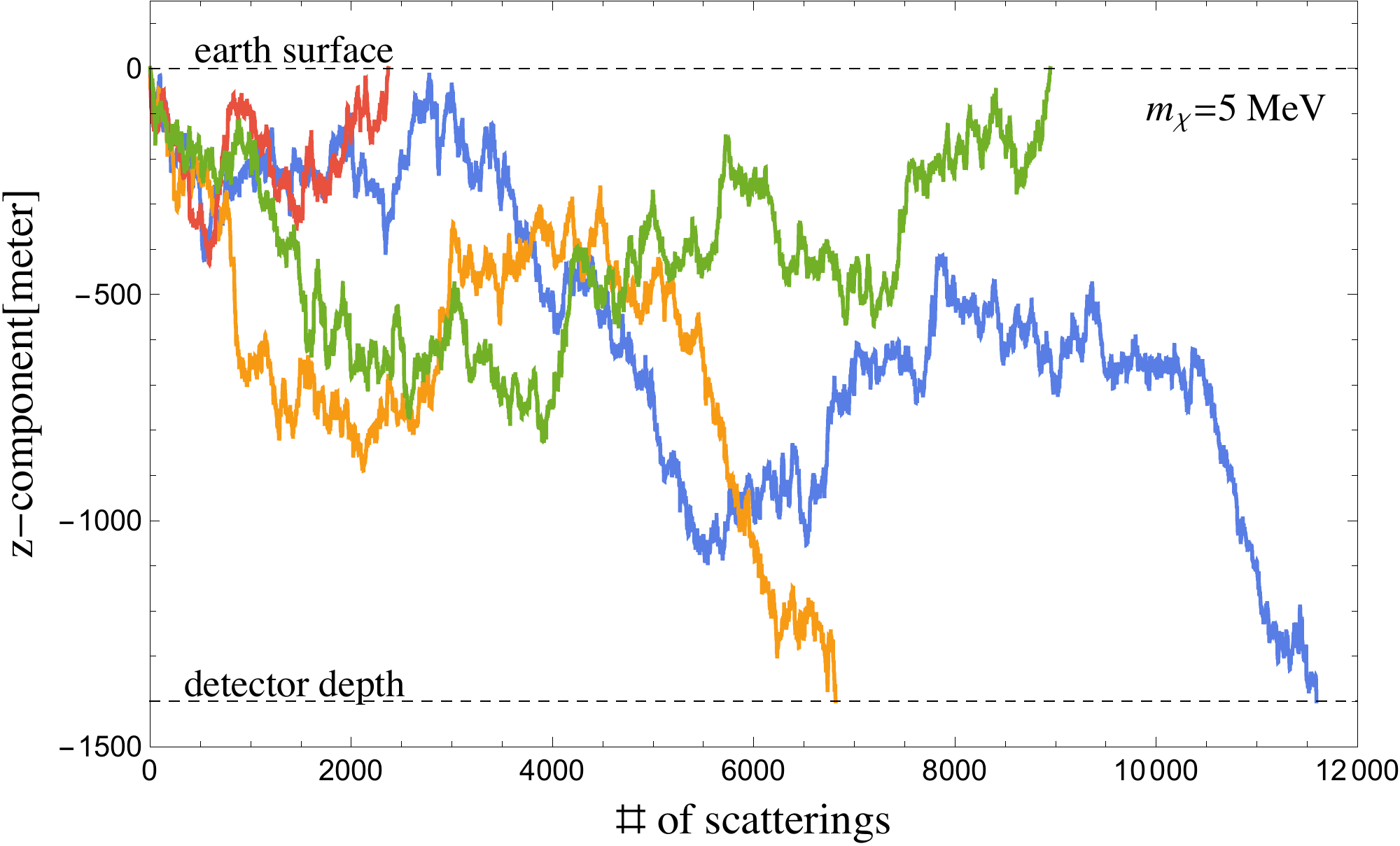}
	\caption{Example trajectories ($z$-component only) as a function of the number of scattering events. The depth is 1400m corresponding to the underground depth of the experimental facilities of the LNGS. The DM-proton scattering cross-section is $\sigma_p\approx 10^{-28}\text{cm}^2$, the critical value for $m_X=5\text{ MeV}$. For further description we refer to the text.}
	\label{fig:trajectories}
\end{figure}
As an illustration we present four example trajectories in Fig. \ref{fig:trajectories}, not as a function of time but as a function of the number of scattering events. Two of these show particles which get reflected and leave the Earth crust again. The other two reach the detector depth of 1400 m after scattering on nuclei thousands of times. The DM-proton scattering cross-section here is given by $\sigma_p\approx 10^{-28}\text{cm}^2$, which following Eq. \eqref{eq:eN} corresponds to the DM-electron scattering cross-section $\sigma_e\approx 10^{-30}\text{cm}^2$. This is the critical value for $m_X=5\text{ MeV}$, above which the screening effect of the Earth crust is strong enough to spoil detectability of DM at the LNGS. The figure illustrates that deflections significantly extend the DM particle's paths between the Earth surface and the detector depth.

\begin{figure}[t!]
	\centering
	\includegraphics[width=0.5\textwidth]{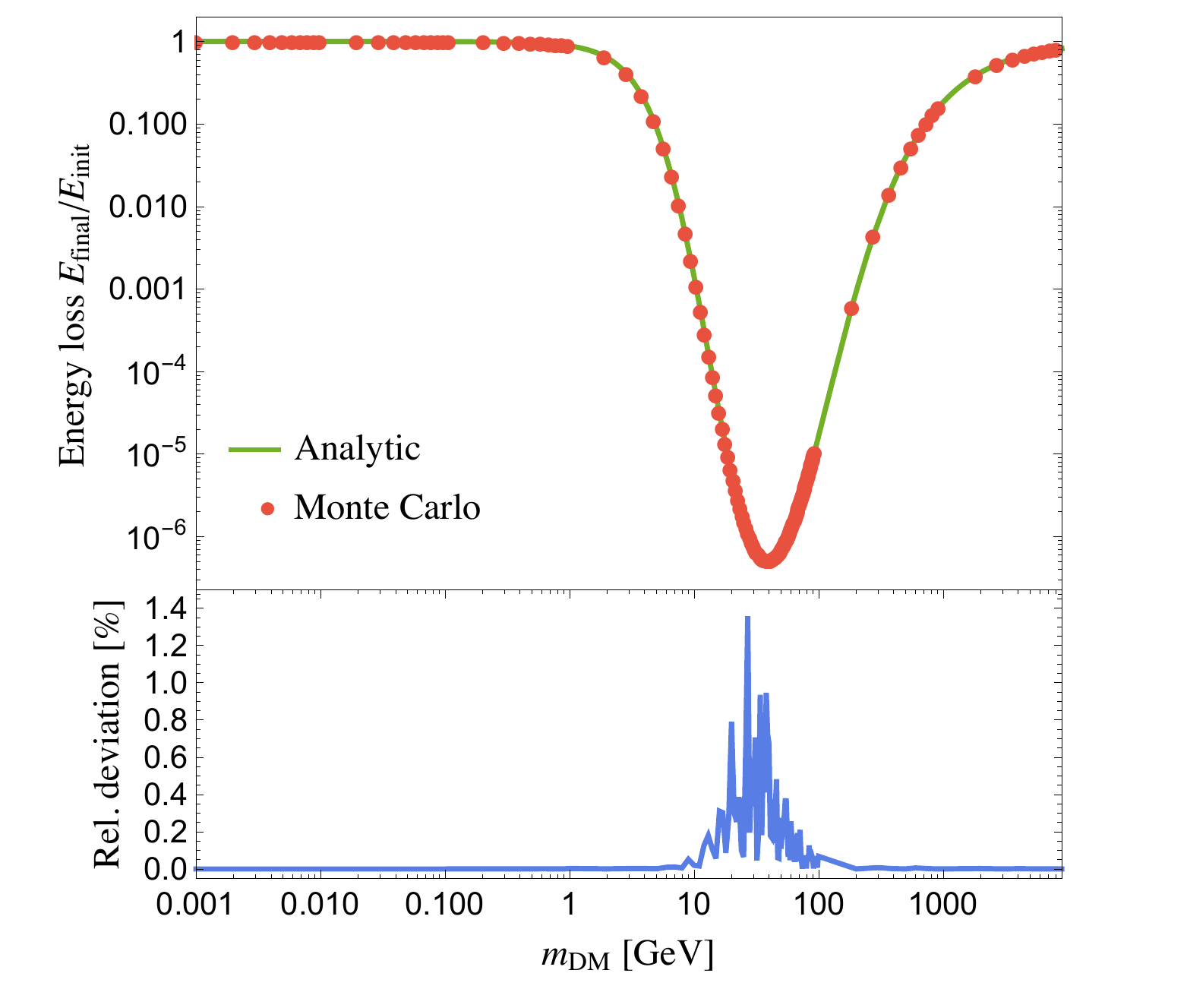}
	\caption{Comparison of the analytical and numerical description of energy loss due to elastic scatterings on oxygen nuclei for $\sigma_p=10^{-32}\text{cm}^2$ and $d=10\text{ km}$. The upper panel shows the average energy loss as a function of $m_X$ in the MC simulations compared to the analytic prediction. The lower panel shows the relative deviations between the two.}
	\label{fig:comparison}
\end{figure}

As a consistency check we would like to compare  the average energy loss estimated using the MC simulation with the analytic expression of Eq. \eqref{eq:cont2}. To make the comparison with Eq. \eqref{eq:cont2} more clear, we only consider oxygen nuclei (which dominate the stopping effect), simulating a large sample of particle trajectories up to a given length $d$ and comparing the average energy loss for different values of $m_X$. The particles' average energy loss in the MC simulations is depicted in the upper panel of Fig. \ref{fig:comparison} along with the analytic prediction. The correspondence is remarkably good over a wide range of masses. The relative deviations are shown in the lower panel. There are some larger deviations for masses of several tens of GeV, where the momentum transfer per scattering and the overall energy loss is maximal. However these deviations are relatively small ($<2\%$) and do not occur for the sub-GeV masses considered in this work. In the relevant mass interval between 1 MeV and 5 GeV the relative deviations never exceed $0.01\%$. However, we should stress here that although the analytic and MC energy losses are practically identical for a given traveled length in the crust of the Earth, they lead eventually to different constraints simply because the MC takes into account the effect of deflection i.e., the DM particle does not move on a straight line during underground collisions. This is something that the analytic formula does not account for because it cannot predict the total length traveled by a particle reaching the depth of the detector if the particle changes path (as it does) every time it scatters.

\subsection{Critical cross-section}
The simulations provide us with a large sample of $N_{\rm tot}$ velocity data points at the detector depth. After the sample size reaches a certain number of particles, which has made it to the detector site, we compute the average velocity and standard deviation,
\begin{align}
	\langle v \rangle &= \frac{1}{N_{\rm tot}}\sum\limits_{i=1}^{N_{\rm tot}} |\vec{v}_i|\, ,\\
	\Delta v &= \sqrt{ \frac{1}{N_{\rm tot}} \sum\limits_{i=1}^{N_{\rm tot}}\left(|\vec{v}_i|-\langle v \rangle\right)^2}\, .
\end{align}
We then perform a parameter scan in $m_{X}$ and $\sigma^{\text{SI}}_p$. For a given mass value $m_{X}$ we find the critical value of the cross-section for which the DM particles are slowed down to velocities below $v_{\text{min}} = \sqrt{\frac{2E_{\rm thr}}{m_X}}$, under which no electron ionisation can occur in the detector. Here we conservatively assume that the DM particle deposits \emph{all} of its kinetic energy into the ionisation process. The critical cross-section is found by systematically increasing the DM-nucleon scattering cross-section, until the mean velocity $\langle v \rangle$ at detector depth is below $v_{\text{min}}$ by five standard deviations, i.e.,
\begin{align}
	\langle v \rangle+ 5\; \Delta v \lesssim v_{\text{min}}\, .
\end{align}

This is our criterion that the DM particles making it to detector depth have been slowed down to non-detectability. Since we are interested in the interaction strength between DM and electrons, not nuclei, we relate the $\sigma^{\rm max}_{p}$ to the DM-electron scattering cross-section $\sigma^{\rm max}_{e}$ via Eq.   \eqref{eq:eN}, just as we did in the analytic treatment.

\section{Results}
\begin{figure}[t!]
\includegraphics[width=.5\textwidth]{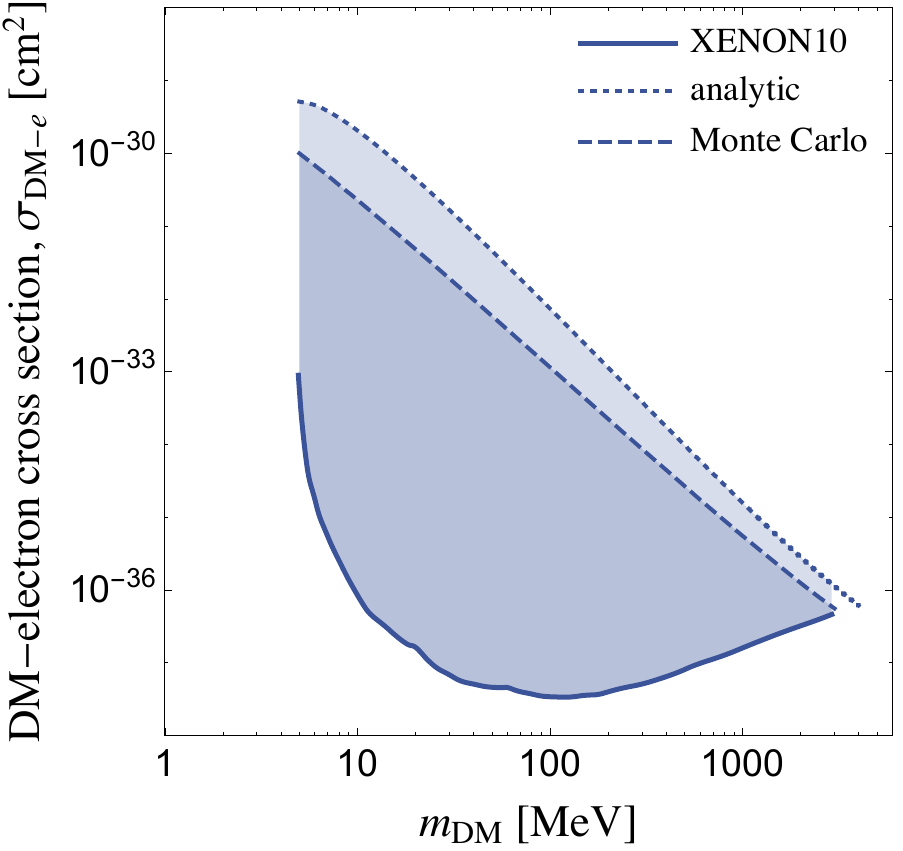}
\includegraphics[width=.5\textwidth]{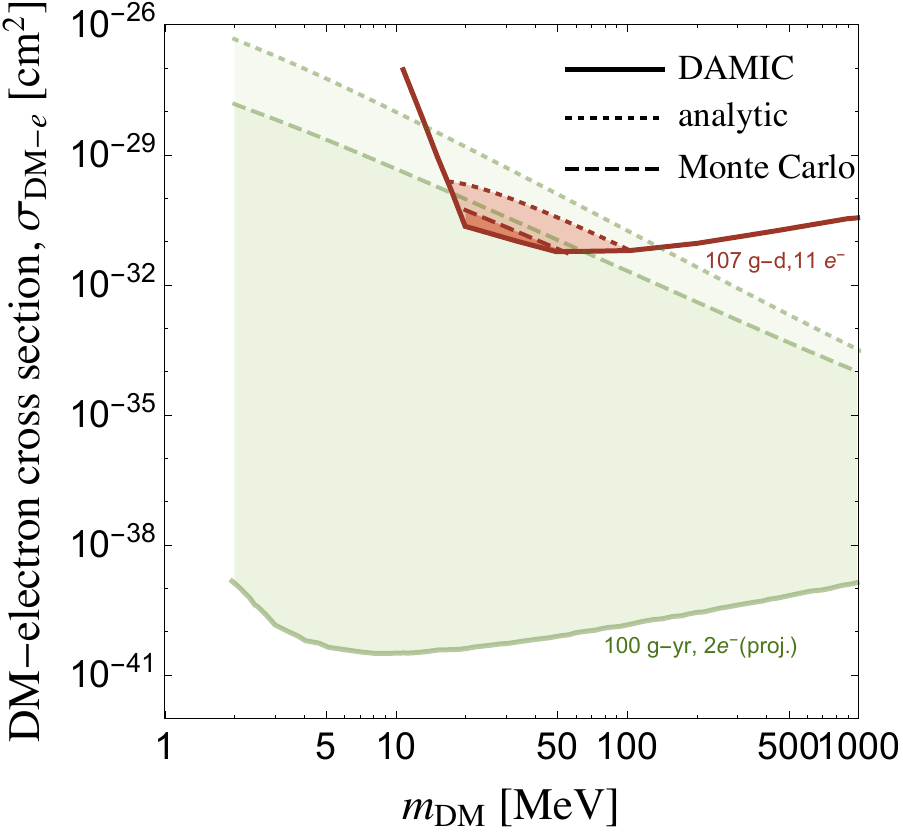}
\caption{XENON and DAMIC constraints including the effects of nuclear stopping. The estimated DAMIC limits for the 2012 data and for the projection have been taken from Fig. 1 of \cite{Essig:2015cda}. {The $11e^-$ and $2e^-$ are the experiment’s thresholds, the minimal number of electron-hole pairs, which lead to a detection. However they also correspond directly to a certain energy thresholds.}}
\label{fig:xenonMC}
\end{figure}

Above a certain DM-nucleus scattering cross-section, incoming DM particles get sufficiently slowed down below the experimental threshold. These cross-section can not be probed in underground facilities. Based on conservative assumptions we employed two methods to find the critical cross-section for this case. For XENON10 and DAMIC the results are depicted in Fig. \ref{fig:xenonMC}.



As expected, the MC estimates are stronger by up to one order of magnitude, and therefore allow less parameter space to be excluded. This is due to the fact that the simulations take deflections into account, which increase the travelled distance of the incoming DM particle and therefore the stopping power of the Earth crust above the experiment's location. {In retrospect the MC results allow us to quantify by how much eq. \eqref{eq:sigmaemax} underestimates this screening effect. The critical cross-section above which nuclear stopping blinds the detector is overestimated, in the case of XENON10 by a factor of $\sim 5$ for $m_{\rm DM}=5$ MeV and $\sim 3$ for $m_{\rm DM}=1$ GeV. For DAMIC the analytic estimate exceeds the MC one by a factor $\sim 9$ for $m_{\rm DM}=1$ MeV and $\sim 3$ for $m_{\rm DM}=1$ GeV respectively. The two estimates approach each other for higher DM masses as expected, as smaller scattering angles in the Earth-frame are favoured for heavier DM and hence Eq.~\eqref{eq:sigmaemax} becomes a better approximation.}

In the upper panel we see the excluded parameter space in the $(m_X,\sigma_e)$ plane for the XENON10 experiment.  Including the nuclear stopping effect drastically reduces the excluded parameter space. The first major observation is that only a DM mass window between 5 MeV and 3 GeV can be constrained. As already mentioned this reduction becomes  more severe for the MC results. 

For the DAMIC experiment we found similar results, which are depicted in the lower panel of Fig. \ref{fig:xenonMC}. The DAMIC limits (plotted in red) were estimated based on the 2012 data of \cite{Essig:2015cda} using an exposure of 107 g-d and a threshold of $\sim40\text{ eV}$ . We also show a projection of DAMIC limits (in light green) from the same authors, assuming an exposure of 100 g-yr and a very low energy threshold of about $\sim 1-2\text{ eV}$. Looking at these limits, it is interesting to note that only a tiny fraction of the claimed excluded parameter space remains excluded, once nuclear stopping is taken into account using MC simulations. Again only a DM mass window can be constrained, which compared to the XENON10 limits is tiny, between 20 and 50 MeV. In addition the excluded DM-electron scattering cross-section for these masses does not even span one order of magnitude. In comparison the projected constraints would still allow to exclude large amounts of the parameter space thanks to the assumed long exposure and extremely low threshold.

{Throughout this work we have focused on the case in which $F_{DM} =1$. However this assumption is invalid if the mediator of DM-electron interactions is light compared to the exchanged momentum. In this case, the DM form factor approaches $F_{DM} \propto1/q^{2}$ in the small mediator mass limit.  Based on the earlier work in~\cite{Lee:2015qva}, the combination of both the screening of the interaction and the requirement that it lead to sizable deflection weakens the effects of stopping. We leave a detailed examination of the validity of neglecting small-angle scattering in the light mediator limit for future work.}

\section{Conclusions}
We have shown that the effect of underground scatterings between DM and nuclei should not be ignored in direct detection experiments sensitive to DM-electron scattering, and we encourage experimental collaborations to include this effect when they compute constraints.  In the case that the interaction comes from the exchange of a dark photon, kinetically mixed with the ordinary photon, the scattering on nuclei prior to arrival at the detector is crucial. We applied both analytic and numerical methods to estimate the energy loss of light DM that could potentially scatter underground and lead to a distortion or in a more extreme case complete absence of the recoil events.  The overall effect of underground scattering of DM is to eliminate the constraints above a critical value of the cross section. This is because cross sections larger than this critical value lead to large energy losses that make DM undetectable in underground detetctors. The numerical MC simulations we performed, provide the DM energy loss which agrees remarkably with the analytic estimate initially presented in~\cite{Kouvaris:2014lpa}. However, the MC simulations lead to a larger stopping effect (thus smaller critical cross section) compared to the analytic prediction, simply because MC simulations take into account deflection of particles unlike the analytic formula.

The stopping effect of the Earth has direct implications for the existing limits on DM-electron scattering from XENON10 and DAMIC. First, these experiments do not simply constrain the cross section from above. Instead at a given dark matter mass they only rule out a band of cross sections: those small enough to be detectable by the time they reach the detector but large enough to be seen in the limited exposure of each experiment. Secondly since the strength of stopping increases as the DM mass increases, we find that such experiments also only constrain a window of DM masses: sufficiently large to produce a detectable recoil (as always) but also light enough that the effects of stopping are not overwhelming. For XENON10 we find that this window of masses is reduced to (5-3000) MeV and for DAMIC even more dramatically reduced to (20-50) MeV. 

\begin{acknowledgments}
The CP$^3$-Origins centre is partially funded by the Danish National Research Foundation, grant number DNRF90.
\end{acknowledgments}
\bibliographystyle{JHEP}

\bibliography{nu}

\end{document}